\documentclass[aps, prl, amsmath, amssymb]{revtex4}

\usepackage{graphics}

\begin{document}

\title{The Gauged Unparticle Action}
\author{A. Lewis Licht \footnote{licht@uic.edu}}
\affiliation{Dept. of Physics\\U. of Illinois at Chicago\\Chicago, 
Illinois 60607}

\begin{abstract}
We show that the unparticle action that is made gauge invariant 
by the inclusion of an open Wilson line factor can be transformed 
into the integral-differential operator action that avoids the use of the 
Wilson line factor.  The two forms of the action should therefore 
give the same Feynman diagrams.  We also show that it is relatively 
easy to construct Feynman diagrams using the operator action.  
\end{abstract}

\maketitle

\section{1. Introduction}\label{S:intro}

One approach to making Georgi's~\cite{Geo-1}~\cite{Geo-2} unparticle action gauge 
invariant has been to 
introduce an open Wilson line factor that obeys the Mandelstam 
condition~\cite{Mand} 
on its derivative~\cite{Tern-1}~\cite{Tern-2}.  Recent 
work~\cite{all-1}~\cite{all-2}~\cite{all-3} has shown that the 
Mandelstam condition 
in this context is mathematically inconsistent.  Another approach to gauging 
the unparticle action has been found~\cite{all-4} that uses an integral-differential operator 
method and avoids the Mandelstam derivative.  In the following we show that 
these two methods are equivalent in that the action using the Wilson line can 
be transformed into the operator action.  They should therefore give the same 
unparticle-gauge field Feynman diagrams.

In Section 2 we review briefly the Wilson line unparticle action.  In Section 3 we 
review the operator action.  In section 4 we show their equivalence. 
The use of the operator action to find gaugeon-unparticle vertexes 
is dicussed In Section 5 and the single gaugeon vertex is derived.  
The double gaugeon vertexes are derived in section 6 and a brief 
discussion of general Gaugeon-Unparticle diagrams is given in section 
7. 

\section{2. The Wilson Line Unparticle Action}

The unparticle action introduced by 
Georgi~\cite{Geo-1}~\cite{Geo-2} has been extended by Terning et al~\cite{Tern-1}~\cite{Tern-2} to 

\begin{equation}\label{E:Kern}
I = \int {d^4 xd^4 y} \Phi _u^\dag  \left( x \right)K\left( {x,y} \right)W_\Lambda  \left( {x,y} \right)\Phi _u \left( y \right)
\end{equation}

where

\begin{equation}
\begin{gathered}
  K\left( {x,y} \right) \doteq \left[ { - \left( {\partial _\mu  \partial ^\mu  } \right)_x  + i\varepsilon } \right]^{2 - d_u } \delta ^4 \left( {x - y} \right) \hfill \\
  \quad \quad \quad  = \frac{{2\sin \left( {\pi d_u } \right)e^{\pi id_u } }}
{{A_{du} }}\int {\frac{{d^4 p}}
{{\left( {2\pi } \right)^4 }}e^{ - ip\left( {x - y} \right)} } \left( {p^2  + i\varepsilon } \right)^{2 - d_u }  \hfill \\ 
\end{gathered} 
\end{equation}

and

\begin{equation}
A_{du}  = \frac{{16\pi ^{5/2} }}
{{\left( {2\pi } \right)^{2d_u } }}\frac{{\Gamma \left( {d_u  + 1/2} \right)}}
{{\Gamma \left( {d_u  - 1} \right)\Gamma \left( {2d_u } \right)}}
\end{equation}

Here $W_{\Lambda}$ is a path ordered Wilson line, involving an integral of the gauge 
field over the path $\Lambda$, introduced to make the action gauge invariant:

\begin{equation}
W_\Lambda  \left( {x,y} \right) = P\exp \left[ { + ig\int_y^x {A_\alpha  \left( \zeta  \right)d\zeta ^\alpha  } } \right]
\end{equation}

The path ordering is from x on the left to y on the right. $W_{\Lambda}$ 
is assumed to satisfy the Mandelstam~\cite{Mand} condition on its derivative,

\begin{equation}\label{E:MC}
\frac{\partial }
{{\partial x^\nu  }}W_\Lambda  \left( {x,y} \right) =  + igA_\nu  \left( x \right)W_\Lambda  \left( {x,y} \right)
\end{equation}

In Ref.~\cite{all-3} it was shown that for an actual Wilson line 
there should be an extra term on the right hand side of Eq.(~\ref{E:MC}).  In 
the following however it is shown that if one assumes there does 
exist some sort of factor that obeys this Mandelstam condition, 
then the Unparticle action can be transformed into a form that does 
not require the $W_{\Lambda}$.

\section{3.  The Operator Unparticle Action}

We review here the differential-integral operator notation for 
action integrals introduced in Ref.~\cite{all-4}.

We introduce kets $\left| {x,i} \right\rangle $that are the eigenkets 
of an abstract position operator $X^\mu  $,  and also carry the gauge 
representation index i,  

satisfying

\begin{equation}
\begin{gathered}
  X^\mu  \left| {x,i} \right\rangle  = x^\mu  \left| {x,i} \right\rangle  \\ 
  \left\langle {x,i} \right|\left. {x',j} \right\rangle  = \delta ^4 \left( {x - x'} \right)\delta _{ij}  \\ 
\end{gathered} 
\end{equation}

and kets $\left| {p,i} \right\rangle $, eigenkets of the conjugate operator
$P^\mu  $,

\begin{equation}
\begin{gathered}
  P_\nu  \left| {p_\nu  ,i} \right\rangle  = p_\nu  \left| {p,i} \right\rangle  \\ 
  \left\langle {p,i} \right|\left. {p',j} \right\rangle  = \delta ^4 \left( {p - p'} \right)\delta _{ij}  \\ 
  \left\langle {x,i} \right|\left. {p,j} \right\rangle  = \frac{{e^{ - ipx} }}
{{\left( {2\pi } \right)^2 }}\delta _{ij}  \\ 
\end{gathered} 
\end{equation}

with

\begin{equation}
\left[ {X^\mu  ,P_\nu  } \right] =  - i\delta _\nu ^\mu  
\end{equation}

We identify the unparticle fields $\Phi _u $with Hilbert space vectors

\begin{equation}
\left| {\Phi _u } \right\rangle  = \int {d^4 x} \left| {x,i} \right\rangle \Phi _u ^i \left( x \right)
\end{equation}

Note that it follows from the above that 

\begin{equation}\label{E:PFI}
\begin{gathered}
  P_\mu  \left| {x,i} \right\rangle  =  - i\partial _\mu  \left| {x,i} \right\rangle  \hfill \\
  P_\mu  \left| {\Phi _u } \right\rangle  = \int {d^4 x} \left| {x,i} \right\rangle i\partial _\mu  \Phi _{_u }^i \left( x \right) \hfill \\
  \left\langle {\Phi _u } \right|P_\mu   = \int {d^4 x} \Phi _\mu ^{i\dag } \left( x \right)\left( { - i\overset{\lower0.5em\hbox{$\smash{\scriptscriptstyle\leftarrow}$}}{\partial } _\mu  } \right)\left\langle {x,i} \right| \hfill \\ 
\end{gathered} 
\end{equation}

Using the branch integral formula

\begin{equation}\label{E:BI}
z^n  =  - \frac{{e^{\pi in} }}
{\pi }\sin \left( {\pi n} \right)\int_0^\infty  {dx\frac{{x^n }}
{{x - z}}} 
\end{equation}

we can write the ungauged unparticle action as

\begin{equation}
I = \left\langle {\Phi _u \left| K \right|\Phi _u } \right\rangle 
\end{equation}

where

\begin{equation}
K = K_0 \int_0^\infty  {dM^2 \frac{{\left( {M^2 } \right)^{2 - d_u } }}
{{M^2  - P^2  - i\varepsilon }}} 
\end{equation}

with

\begin{equation}
K_0  = \frac{{2\sin ^2 \left( {\pi d_u } \right)}}
{{\pi A_{du} }}
\end{equation}

The gauge field $A_\mu  $can be considered as an operator on this Hilbert space, with

\begin{equation}
A_\mu   = \int {d^4 x} \left| {x,i} \right\rangle A_\mu ^a \left( x \right)T_{ij}^a \left\langle {x,j} \right|
\end{equation}

Under the gauge transform

\begin{equation}
\left| {\Phi _u } \right\rangle  \to \left| {\Phi _u '} \right\rangle  = e^{ + ig\Lambda } \left| {\Phi _u } \right\rangle 
\end{equation}

\begin{equation}
A_\mu   \to A_\mu  ' = e^{ + ig\Lambda } A_\mu  e^{ - ig\Lambda }  + \frac{1}
{g}\left[ {e^{ + ig\Lambda } ,P_\mu  } \right]e^{ - ig\Lambda } 
\end{equation}

The combination

\begin{equation}
D_\mu   = P_\mu   + gA_\mu  
\end{equation}

transforms as

\begin{equation}
D_\mu   \to D_\mu  ' = e^{ + ig\Lambda } D_\mu  e^{ - ig\Lambda } 
\end{equation}

so that replacing $P_{\mu}$ by $D_{\mu}$ gives us the gauge invariant action,

\begin{equation}\label{E:GIA}
\begin{gathered}
  I = K_0 \int_0^\infty  {dM^2 } \left( {M^2 } \right)^{2 - d_u } \left\langle {\Phi _u } \right|\frac{1}
{{M^2  - D^2  - i\varepsilon }}\left| {\Phi _u } \right\rangle  \\ 
   = K_0 \int_0^\infty  {dM^2 } \left( {M^2 } \right)^{2 - d_u } \left\langle {\Phi _u } \right|\frac{1}
{{M^2  - P^2  - g\left\{ {P^\mu  ,A_\mu  } \right\} - g^2 A_\mu  A^\mu   - i\varepsilon }}\left| {\Phi _u } \right\rangle  \\ 
\end{gathered} 
\end{equation}

\section{4.  The Equivalence Between the Actions.}

We will assume here that there actually is a factor W(x,y) such that

\begin{equation}\label{E:ASMW}
\frac{\partial }
{{\partial x^\mu  }}W\left( {x,y} \right) =  + igA_\mu  \left( x \right)W\left( {x,y} \right)
\end{equation}

and that

\begin{equation}\label{E:W0}
W\left( {x,x} \right) = 1
\end{equation}

Using the branch cut integral of Eq. (~\ref{E:BI}) ,  the kernel of 
Eq. (~\ref{E:Kern}) can be written as 

\begin{equation}
\begin{gathered}
  K\left( {x,y} \right) =  = K_0 \int_0^\infty  {dM^2 \frac{{\left( {M^2 } \right)^{2 - d_u } }}
{{M^2  + \partial _x^2  - i\varepsilon }}} \int {\frac{{d^4 p}}
{{\left( {2\pi } \right)^4 }}e^{ - ip\left( {x - y} \right)} }  \\ 
   = K_0 \int_0^\infty  {dM^2 \frac{{\left( {M^2 } \right)^{2 - d_u } }}
{{M^2  + \partial _x^2  - i\varepsilon }}} \delta ^4 \left( {x - y} \right) \\ 
   = \sum\limits_{n = 0}^\infty  {\left( { - 1} \right)^n E_n } \left( {\partial _{x\mu } \partial _x^\mu   - i\varepsilon } \right)^n \delta ^4 \left( {x - y} \right) \\ 
\end{gathered} 
\end{equation}

with the parameters $E_n $ given by

\begin{equation}
E_n  = Lim_{m \to 0}K_{0} \int_{m^2 }^\infty  {dM^2 } \left( {M^2 } \right)^{1 - d_u  - n} 
\end{equation}

where the limit is to be taken after the summation.

Now, the integral

\begin{equation}
\begin{gathered}
  I_n  = \left( { - 1} \right)^n E_n \int {d^4 xd^4 y} \Phi _u^\dag  \left( x \right)\left( {\partial _{x\mu } \partial _x^\mu   - i\varepsilon } \right)^n \delta ^4 \left( {x - y} \right)W\left( {x,y} \right)\Phi _u \left( y \right) \\ 
   = \left( { - 1} \right)^n E_n \int {d^4 xd^4 y} \Phi _u^\dag  \left( x \right)\left[ {\left( { - \overset{\lower0.5em\hbox{$\smash{\scriptscriptstyle\leftarrow}$}}{\partial } _{x\mu }  - igA_\mu  } \right)\left( { - \overset{\lower0.5em\hbox{$\smash{\scriptscriptstyle\leftarrow}$}}{\partial } _x^\mu   - igA^\mu  } \right) - i\varepsilon } \right]^n \delta ^4 \left( {x - y} \right)W\left( {x,y} \right)\Phi _u \left( y \right) \\ 
\end{gathered} 
\end{equation}

after integrating by parts and using Eq. (~\ref{E:ASMW}).   Integrating out y and using 
Eq. (~\ref{E:W0}) makes this

\begin{equation}
\begin{gathered}
  I_n  = \left( { - 1} \right)^n E_n \int {d^4 x} \Phi _u^\dag  \left( x \right)\left[ {\left( { - \overset{\lower0.5em\hbox{$\smash{\scriptscriptstyle\leftarrow}$}}{\partial } _{x\mu }  - igA_\mu  } \right)\left( { - \overset{\lower0.5em\hbox{$\smash{\scriptscriptstyle\leftarrow}$}}{\partial } _x^\mu   - igA^\mu  } \right) - i\varepsilon } \right]^n \Phi _u \left( x \right) \\ 
   = E_n \int {d^4 x} \Phi _u^\dag  \left( x \right)\left[ {\left( { - i\overset{\lower0.5em\hbox{$\smash{\scriptscriptstyle\leftarrow}$}}{\partial } _{x\mu }  + gA_\mu  } \right)\left( { - i\overset{\lower0.5em\hbox{$\smash{\scriptscriptstyle\leftarrow}$}}{\partial } _x^\mu   + gA^\mu  } \right) + i\varepsilon } \right]^n \Phi _u \left( x \right) \\ 
\end{gathered} 
\end{equation}

Using Eq. (~\ref{E:PFI}) this can be written in the operator notation of Section (3) as

\begin{equation}
\begin{gathered}
  I_n  = E_n \left\langle {\Phi _u } \right|\left[ {\left( {P_\mu   + gA_\mu  } \right)\left( {P_\mu   + gA_\mu  } \right) + i\varepsilon } \right]^n \left| {\Phi _u } \right\rangle  \\ 
   = E_n \left\langle {\Phi _u } \right|\left( {D^2  + i\varepsilon } \right)^n \left| {\Phi _u } \right\rangle  \\ 
\end{gathered} 
\end{equation}

Summing, we get the action of Eq. (~\ref{E:GIA}).

\section{5.  Gaugeon-Unparticle Vertexes}

The operator method gives a relatively simple way of deriving unparticle-gauge 
field vertexes.  Let 

\begin{equation}
G_M \left( P \right) = \frac{1}
{{M^2  - P^2  - i\varepsilon }}
\end{equation}

Then Eq. (~\ref{E:GIA}) can be written as

\begin{equation}\label{E:GZG}
I = \sum\limits_{n = 0}^\infty  {} K_0 \int_0^\infty  {dM^2 } \left( {M^2 } \right)^{2 - d_u } \left\langle {\Phi _u } \right|G_M \left( P \right)\left[ {ZG_M \left( P \right)} \right]^n \left| {\Phi _u } \right\rangle 
\end{equation}

where

\begin{equation}\label{E:ZEQ}
Z = g\left\{ {P^\mu  ,A_\mu  } \right\} + g^2 A_\mu  A^\mu  
\end{equation}

Recall that the propagator for the gauge field in Feynman gauge is 

\begin{equation}
\begin{gathered}
  \left\langle \Omega  \right|TA_\alpha ^a \left( x \right)A_\beta ^b \left( y \right)\left| \Omega  \right\rangle _0  = iD_A \left( {x - y} \right)\eta _{\alpha \beta } \delta ^{ab}  \\ 
   =  - \eta _{\alpha \beta } \delta ^{ab} \int {d^4 pe^{ - ip \cdot \left( {x - y} \right)} } \frac{{ - i}}
{{\left( {2\pi } \right)^4 }}\frac{1}
{{ - p^2  - i\varepsilon }} \\ 
   = i\eta _{\alpha \beta } \delta ^{ab} \int {d^4 pe^{ - ip \cdot \left( {x - y} \right)} } \frac{1}
{{\left( {2\pi } \right)^4 }}\tilde D_A \left( p \right) \\ 
\end{gathered} 
\end{equation}

Where we denote the particle vacuum state by $\Omega$ and the zeroth order 
propagator by a zero subscript. 

The propagator for the unparticle is

\begin{equation}
\begin{gathered}
  \left\langle \Omega  \right|T\Phi _u^i \left( x \right)\Phi _u^{j\dag } \left( y \right)\left| \Omega  \right\rangle _0  = i\delta ^{ij} S\left( {x - y} \right) \\ 
   = i\delta ^{ij} \int {d^4 p} e^{ - ip \cdot \left( {x - y} \right)} \frac{1}
{{\left( {2\pi } \right)^4 }}\tilde S\left( p \right) \\ 
   = i\delta ^{ij} \int {d^4 p} e^{ - ip \cdot \left( {x - y} \right)} \frac{1}
{{\left( {2\pi } \right)^4 }}\int_0^\infty  {dM^2 \left( {M^2 } \right)^{d_u  - 2} } \frac{{ - iA_{du} }}
{{2\pi }}G_M \left( p \right) \\ 
\end{gathered} 
\end{equation}

We will use here the notation

\begin{equation}\label{E:FIFTW}
\begin{gathered}
  \Phi _u \left( x \right) = \int {\frac{{d^4 p}}
{{\left( {2\pi } \right)^4 }}} e^{ - ip \cdot x} \tilde \Phi _u \left( p \right) \\ 
  A_\alpha ^a \left( x \right) = \int {\frac{{d^4 p}}
{{\left( {2\pi } \right)^4 }}} e^{ - ip \cdot x} \tilde A_\alpha ^a \left( p \right) \\ 
\end{gathered} 
\end{equation}

Then

\begin{equation}
\begin{gathered}
  \left| {\Phi _u } \right\rangle  = \int {d^4 x} \left| {x,i} \right\rangle \Phi _u^i \left( x \right) \\ 
   = \int {\frac{{d^4 p}}
{{\left( {2\pi } \right)^2 }}} \left| {p,i} \right\rangle \tilde \Phi _u^i \left( p \right) \\ 
\end{gathered} 
\end{equation}

and

\begin{equation}
\begin{gathered}
  A_\alpha   = \int {d^4 x} \left| {x,i} \right\rangle A_\alpha ^a \left( x \right)T_{ij}^a \left\langle {x,i} \right| \\ 
   = \int {\frac{{d^4 qd^4 k}}
{{\left( {2\pi } \right)^4 }}} \left| {q + k,i} \right\rangle \tilde A_\alpha ^a \left( q \right)T_{ij}^a \left\langle {k,i} \right| \\ 
\end{gathered} 
\end{equation}

We consider first the single gaugeon vertex.  The single gauge field 
interaction can be seen from Eqs. (~\ref{E:GZG}) and (~\ref{E:ZEQ}) to be 

\begin{equation}
I_1  = gK_0 \int_0^\infty  {dM^2 } \left( {M^2 } \right)^{2 - d_u } \left\langle {\Phi _u } \right|G_M \left( P \right)\left\{ {P^\mu  ,A_\mu  } \right\}G_M \left( P \right)\left| {\Phi _u } \right\rangle 
\end{equation}

Which can be written now as

\begin{equation}\label{E:I1}
\begin{gathered}
  I_1  = gK_0 \int_0^\infty  {dM^2 } \left( {M^2 } \right)^{2 - d_u } \int {\frac{{d^4 p'd^4 qd^4 \ell d^4 p}}
{{\left( {2\pi } \right)^8 }}} \tilde \Phi _u^\dag  \left( {p'} \right)\left\langle {p'} \right|G_M \left( {p'} \right)\left| {q + \ell } \right\rangle \left( {q + 2\ell } \right)^\mu   \\ 
  \quad \quad \tilde A_\mu  \left( q \right)\left\langle \ell  \right|G_M \left( p \right)\left| p \right\rangle \tilde \Phi _u \left( p \right) \\ 
   = gK_0 \int_0^\infty  {dM^2 } \left( {M^2 } \right)^{2 - d_u } \int {\frac{{d^4 p'd^4 qd^4 p}}
{{\left( {2\pi } \right)^8 }}\delta \left( {p' - q - p} \right)}  \\ 
  \quad \quad \tilde \Phi _u^\dag  \left( {p'} \right)G_M \left( {p'} \right)\left( {p' + p} \right)^\mu  \tilde A_\mu  \left( q \right)G_M \left( p \right)\tilde \Phi _u \left( p \right) \\ 
\end{gathered} 
\end{equation}

Where we have suppressed the representation indices.  

We consider now the vacuum expectation value of the Heisenberg fields:

\begin{equation}
V_\alpha ^{ija} \left( {u,v,w} \right) = \left\langle \Omega  \right|T\left( {\Phi _{_u }^i \left( u \right)A_\alpha ^a \left( v \right)\Phi _u^{j\dag } \left( w \right)} \right)\left| \Omega  \right\rangle 
\end{equation}

To lowest order in g, this would be given by

\begin{equation}
V_\alpha ^{ija} \left( {u,v,w} \right) = i^3 \int {d^4 u'd^4 v'd^4 w'} S\left( {u - u'} \right)D_A \left( {v' - v} \right)S\left( {w' - w} \right)\frac{\delta }
{{\delta \Phi _u^{i\dag } \left( {u'} \right)}}\frac{\delta }
{{\delta A^{a\alpha } \left( {v'} \right)}}\frac{\delta }
{{\delta \Phi _u^j \left( {w'} \right)}}iI_1 
\end{equation}

Now inverting Eq. (~\ref{E:FIFTW}), we have

\begin{equation}
\begin{gathered}
  \tilde \Phi _u \left( p \right) = \int {d^4 x} e^{ip \cdot x} \Phi _u \left( x \right) \\ 
  \tilde A_\alpha ^a \left( p \right) = \int {d^4 x} e^{ip \cdot x} A_\alpha ^a \left( x \right) \\ 
\end{gathered} 
\end{equation}

From which we get,

\begin{equation}
\begin{gathered}
  \frac{\delta }
{{\delta \Phi _u^i \left( x \right)}} = \int {d^4 p} \frac{{\delta \tilde \Phi _u^i \left( p \right)}}
{{\delta \Phi _u^i \left( x \right)}}\frac{\delta }
{{\delta \tilde \Phi _u^i \left( p \right)}} \\ 
   = \int {d^4 p} e^{ip \cdot x} \frac{\delta }
{{\delta \tilde \Phi _u^i \left( p \right)}} \\ 
\end{gathered} 
\end{equation}

and similiarly,

\begin{equation}
\frac{\delta }
{{\delta A_\alpha ^a \left( x \right)}} = \int {d^4 p} e^{ip \cdot x} \frac{\delta }
{{\delta \tilde A_\alpha ^a \left( p \right)}}
\end{equation}

Equivalently,

\begin{equation}\label{E:DTW}
\begin{gathered}
  \frac{\delta }
{{\delta \tilde \Phi _u^i \left( p \right)}} = \int {d^4 x} \frac{{\delta \Phi _u^i \left( x \right)}}
{{\delta \tilde \Phi _u^i \left( p \right)}}\frac{\delta }
{{\delta \Phi _u^i \left( x \right)}} \\ 
   = \int {\frac{{d^4 x}}
{{\left( {2\pi } \right)^4 }}} e^{ - ip \cdot x} \frac{\delta }
{{\delta \Phi _u^i \left( x \right)}} \\ 
  \frac{\delta }
{{\delta \tilde A_\alpha ^a \left( p \right)}} = \int {\frac{{d^4 x}}
{{\left( {2\pi } \right)^4 }}} e^{ - ip \cdot x} \frac{\delta }
{{\delta A_\alpha ^a \left( x \right)}} \\ 
\end{gathered} 
\end{equation}

The vacuum expectation value in momentum space is

\begin{equation}
\tilde V_\alpha ^{ija} \left( {p',q,p} \right) = \int {d^4 ud^4 vd^4 w} e^{i\left( {p' \cdot u - q \cdot v - p \cdot w} \right)} V_\alpha ^{ija} \left( {u,v,w} \right)
\end{equation}

Which becomes

\begin{equation}
\tilde V_\alpha ^{ija} \left( {p',q,p} \right) = i^3 \int {d^4 u'd^4 v'd^4 w'} \tilde S\left( {p'} \right)\tilde D_A \left( q \right)\tilde S\left( p \right)e^{i\left( {p' \cdot u' - q \cdot v' - p \cdot w'} \right)} \frac{\delta }
{{\delta \Phi _u^{i\dag } \left( {u'} \right)}}\frac{\delta }
{{\delta A^{\alpha a} \left( {v'} \right)}}\frac{\delta }
{{\delta \Phi _u^j \left( {w'} \right)}}iI_1 
\end{equation}

Eq. (~\ref{E:DTW}) makes this into

\begin{equation}
\tilde V_\alpha ^{ija} \left( {p',q,p} \right) = i^3 \tilde S\left( {p'} \right)\tilde D_A \left( q \right)\tilde S\left( p \right)\left( {2\pi } \right)^{12} \frac{\delta }
{{\delta \tilde \Phi _u^{i\dag } \left( {p'} \right)}}\frac{\delta }
{{\delta \tilde A^{\alpha a} \left( q \right)}}\frac{\delta }
{{\delta \Phi _u^j \left( p \right)}}iI_1 
\end{equation}

The vertex function is defined as 

\begin{equation}
\tilde V_\alpha ^{ija} \left( {p',q,p} \right) = i^3 \tilde S\left( {p'} \right)\tilde D_A \left( q \right)\tilde S\left( p \right)ig\left( {2\pi } \right)^4 \delta \left( {p' - q - p} \right)\Gamma _\alpha ^{ija} \left( {p',q,p} \right)
\end{equation}

The one gaugeon vertex is thus

\begin{equation}
ig\left( {2\pi } \right)^4 \delta \left( {p' - q - p} \right)\Gamma _\alpha ^{ija} \left( {p',q,p} \right) = \left( {2\pi } \right)^{12} \frac{\delta }
{{\delta \tilde \Phi _u^{i\dag } \left( {p'} \right)}}\frac{\delta }
{{\delta \tilde A^{\alpha a} \left( q \right)}}\frac{\delta }
{{\delta \Phi _u^j \left( p \right)}}iI_1 
\end{equation}

Which becomes, using Eq. (~\ref{E:I1}),

\begin{equation}
\Gamma ^{ij\alpha a} \left( {p',q,p} \right) = K_0 \left( {p' + p} \right)^\alpha  T_{ij}^a \int_0^\infty  {dM^2 } \left( {M^2 } \right)^{2 - d_u } G_M \left( {p'} \right)G_M \left( p \right)
\end{equation}

It can easily be shown that 

\begin{equation}\label{E:GPG}
G_M \left( {p'} \right)G_M \left( p \right) = \frac{1}
{{p'^2  - p^2 }}\left[ {G_M \left( {p'} \right) - G_M \left( p \right)} \right]
\end{equation}

which makes the vertex into

\begin{equation}
\Gamma ^{ij\alpha a} \left( {p',q,p} \right) = \frac{{\left( {p' + p} \right)^\alpha  }}
{{p'^2  - p^2 }}T_{ij}^a \left[ {\tilde S^{ - 1} \left( {p'} \right) - \tilde S^{ - 1} \left( p \right)} \right]
\end{equation}

Exactly the form found in Ref.~\cite{Tern-1}.

\section{6.  The Two Gaugeon Vertexes.}

We consider here the general case of an n gaugeon vertex, and give specific 
results for the n= 2 case. 
 
Eq. (~\ref{E:GPG}) is a special case of a general result that is useful for the multiple 
gaugeon vertexes. 

\begin{equation}\label{E:MGV}
\prod\nolimits_{k = 1}^n {\frac{1}
{{M^2  - p_k^2  - i\varepsilon }}}  = \sum\nolimits_{k = 1}^n {\frac{1}
{{M^2  - p_k^2  - i\varepsilon }}\prod\nolimits_{j \ne k} {\frac{1}
{{p_k^2  - p_j^2 }}} } 
\end{equation}

The left hand side of this equation is an analytic function 
$F_L \left( {M^2 } \right)$ with poles at the points $M^2  = p_k^2 $ and the 
right hand side is another analytic function $F_R \left( {M^2 } \right)$ with 
poles at the same points, and with the same residues.  The difference, 
$F\left( {M^2 } \right) = F_L \left( {M^2 } \right) - F_R \left( {M^2 } \right)$
is an analytic function with no poles, and goes to zero as  $M^2 $
goes to infinity.  It is therefore zero by LiouvilleÕs 
theorem~\cite{W&W}, which proves Eq. (~\ref{E:MGV}). 

The n-gaugeon vertex can be expressed as

\begin{equation}
\begin{gathered}
  ig^n \left( {2\pi } \right)^4 \delta \left( {p' - \sum {q_k  - p} } \right)\Gamma _{\alpha _1  \cdots \alpha _n }^{ija_1  \cdots a_n } \left( {p',q_1  \cdots q_n ,p} \right) =  \hfill \\
  \quad \quad \quad \quad \quad \quad \quad \left( {2\pi } \right)^{4\left( {n + 2} \right)} \frac{\delta }
{{\delta \tilde \Phi _u^{i\dag } \left( {p'} \right)}}\left( {\prod\limits_{k = 1}^n {} \frac{\delta }
{{\delta \tilde A^{\alpha _k a_k } \left( {q_k } \right)}}} \right)\frac{\delta }
{{\delta \Phi _u^j \left( p \right)}}iI_n  \hfill \\ 
\end{gathered} 
\end{equation}

which yields an expression of the form

\begin{equation}
\Gamma _{\alpha _1  \cdots \alpha _n }^{ija_1  \cdots a_n } \left( {p',q_1  \cdots q_n ,p} \right) = K_0 \int_0^\infty  {dM^2 } \left( {M^2 } \right)^{2 - d_u } \left\{ {\sum {} G_M \left( {p'} \right)\prod {\left( {ZG_M \left( k \right)} \right)} } \right\}_n 
\end{equation}

where the sum is over the various permutations of the external gaugeon 
lines and the term with n total gaugeon lines is to be taken.

For example, we give the explicit expansion for n = 2.

\begin{equation}
\begin{gathered}
  \Gamma ^{ab\alpha \beta } \left( {p',q',q,p} \right) = K_0 \int_0^\infty  {dM^2 } \left( {M^2 } \right)^{2 - d_u } G_M \left( {p'} \right)\left\{ {G_M \left( {p + q} \right)\left( {p' + p + q} \right)^\alpha  \left( {2p + q} \right)^\beta  T^a T^b } \right. \hfill \\
  \quad \quad \quad \quad \quad \quad \quad \quad  + \left. {G_M \left( {p + q'} \right)\left( {p' + p + q'} \right)^\beta  \left( {2p + q'} \right)^\alpha  T^b T^a  + \left\{ {T^a ,T^b } \right\}\eta ^{\alpha \beta } } \right\}^{ij} G_M \left( p \right) \hfill \\ 
\end{gathered} 
\end{equation}

Using Eq. (~\ref{E:MGV}) this can be expressed as

\begin{equation}
\begin{gathered}
  \Gamma ^{ab\alpha \beta } \left( {p',q',q,p} \right) = \left( {p' + p + q} \right)^\alpha  \left( {2p + q} \right)^\beta  T^a T^b \left[ {\frac{{S^{ - 1} \left( {p'} \right)}}
{{\left( {p'^2  - \left( {p + q} \right)^2 } \right)\left( {p'^2  - p^2 } \right)}}} \right. \hfill \\
  \quad \quad \quad \quad \quad \quad \quad  + \left. {\frac{{S^{ - 1} \left( p \right)}}
{{\left( {p^2  - \left( {p + q} \right)^2 } \right)\left( {p^2  - p'^2 } \right)}} + \frac{{S^{ - 1} \left( {p + q} \right)}}
{{\left( {p^2  - \left( {p + q} \right)^2 } \right)\left( {p'^2  - \left( {p + q} \right)^2 } \right)}}} \right] \hfill \\
  \quad \quad \quad \quad \quad \quad \quad  + \left( {p' + p + q'} \right)^\beta  \left( {2p + q'} \right)^\alpha  T^b T^a \left[ {\frac{{S^{ - 1} \left( {p'} \right)}}
{{\left( {p'^2  - \left( {p + q'} \right)^2 } \right)\left( {p'^2  - p^2 } \right)}}} \right. \hfill \\
  \quad \quad \quad \quad \quad \quad \quad  + \left. {\frac{{S^{ - 1} \left( p \right)}}
{{\left( {p^2  - \left( {p + q'} \right)^2 } \right)\left( {p^2  - p'^2 } \right)}} + \frac{{S^{ - 1} \left( {p + q'} \right)}}
{{\left( {p^2  - \left( {p + q'} \right)^2 } \right)\left( {p'^2  - \left( {p + q'} \right)^2 } \right)}}} \right] \hfill \\
  \quad \quad \quad \quad \quad \quad \quad  + \left\{ {T^a ,T^b } \right\}\eta ^{\alpha \beta } \frac{{S^{ - 1} \left( {p'} \right) - S^{ - 1} \left( p \right)}}
{{p'^2  - p^2 }} \hfill \\
  \quad \quad \quad \quad  \hfill \\ 
\end{gathered} 
\end{equation}

These vertexes have also been derived using the Terning method by 
Liao~\cite{Liao-1}~\cite{Liao-2}.

\section{7.  Diagrams.}

We will use diagrams to express these gaugeon-unparticle vertexes in which 
the integral over the unparticle mass is indicated by a heavy line extending 
between two small circles.  The gaugeons will be indicated by dashed lines.  
The segments of the unparticle line between intersections with gaugeon lines 
correspond to $G_M $ factors.  The single gaugeon vertex is then shown in Figure 1.  
The two main contributions to the double gaugeon vertexes are shown in Figures 2 
and 3.  A third contribution to the unparticle-double gaugeon vertex, 
$\Gamma_{2}$, that comes from two single gaugeon vertexes connected by an unparticle 
propagator is shown in Figure 4. Reading it off from the figure and 
symmetrising with respect to the gaugeons gives

\begin{equation}
\begin{gathered}
  \Gamma _2^{ab\alpha \beta } \left( {p',q',q,p} \right) =  - \frac{{\left( {p' + p + q} \right)^\alpha  }}
{{p'^2  - \left( {p + q} \right)^2 }}T^a \left[ {S^{ - 1} \left( {p'} \right) - S^{ - 1} \left( {p + q} \right)} \right]S\left( {p + q} \right) \hfill \\
  \quad \quad \quad \quad \quad \quad \quad \quad  \times \frac{{\left( {2p + q} \right)^\beta  }}
{{\left( {p + q} \right)^2  - p^2 }}T^b \left[ {S^{ - 1} \left( {p + q} \right) - S^{ - 1} \left( p \right)} \right] \hfill \\
  \quad \quad \quad \quad \quad \quad \quad \quad  - \frac{{\left( {p' + p + q'} \right)^\beta  }}
{{p'^2  - \left( {p + q'} \right)^2 }}T^b \left[ {S^{ - 1} \left( {p'} \right) - S^{ - 1} \left( {p + q'} \right)} \right]S\left( {p + q'} \right) \hfill \\
  \quad \quad \quad \quad \quad \quad \quad \quad  \times \frac{{\left( {2p + q'} \right)^\alpha  }}
{{\left( {p + q'} \right)^2  - p^2 }}T^a \left[ {S^{ - 1} \left( {p + q'} \right) - S^{ - 1} \left( p \right)} \right] \hfill \\ 
\end{gathered} 
\end{equation}

\begin{figure}[ht]
    \includegraphics{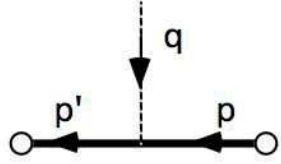}
    \caption{The one gaugeon vertex}
\end{figure}

\begin{figure}[ht]
    \includegraphics{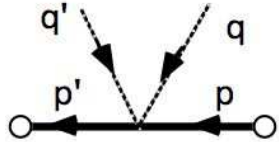}
    \caption{The $A^2 $contribution to the double gaugeon-unparticle vertex.}
\end{figure}

\begin{figure}[ht]
    \includegraphics{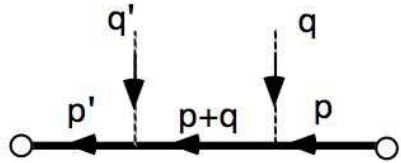}
    \caption{The $\left\{ {P,A} \right\}G_M \left\{ {P,A} \right\}$ contribution 
    to the double gaugeon vertex.}
\end{figure}

\begin{figure}[ht]
    \includegraphics{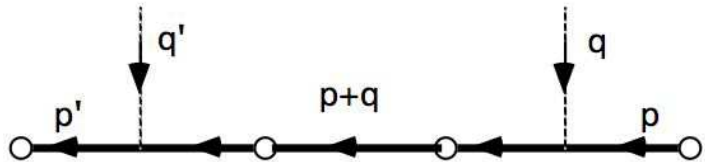}
    \caption{The remaining contribution to the double gaugeon vertex.}
\end{figure}

\section{8. Conclusions}

We have shown that the bosonic unparticle action made gauge invariant by an 
open Wilson line factor can be transformed into the differential-integral 
operator unparticle action.  The two actions should therefore give 
the same vertexes. This is done by ignoring, as is usually done, the extra term that 
should be present on the right hand side of Eq. 
(~\ref{E:MC}).~\cite{all-3} It is 
conceivable that the vertexes would be different if this extra term 
was actually taken into account.

We have also shown that there is a diagramatic technique that allows 
one to read off fairly easily the unparticle-gaugeon vertexes from the differential-integral 
operator unparticle action.

\end{document}